\documentclass{article}

\usepackage[dblblindworkshop,final]{neurips_2026}
\workshoptitle{Interpreting Agent Behavior}

\usepackage[T1]{fontenc}
\usepackage{textcomp}
\usepackage{amsmath}
\usepackage{amssymb}
\usepackage[hidelinks]{hyperref}
\usepackage{url}

\usepackage{booktabs}
\usepackage{tabularx}
\usepackage{array}
\usepackage{graphicx}
\usepackage[expansion=false]{microtype}
\usepackage{xcolor}
\usepackage{enumitem}
\usepackage{tikz}
\usetikzlibrary{arrows.meta,positioning}

\definecolor{barc}{RGB}{88,104,150}
\newlength{\barunit}
\newcommand{\pbar}[1]{{\color{barc}\rule{#1\barunit}{1.35ex}}}
\newcolumntype{L}{>{\raggedright\arraybackslash}X}

\newcommand{\bench}{sudo~L7}
\AtBeginDocument{\setcitestyle{semicolon}}
\newcommand{\doi}[1]{doi:~\href{https://doi.org/#1}{\nolinkurl{#1}}}

\newcommand{\NCriteria}{1{,}287}
\newcommand{\CritMedian}{22}
\newcommand{\CritMin}{7}
\newcommand{\CritMax}{57}
\newcommand{\NCruxTasks}{54}
\newcommand{\NCheckTasks}{44}
\newcommand{\NRepos}{25}
\newcommand{\NBrowserTasks}{17}
\newcommand{\NGating}{374}

\title{sudo L7: Can Coding Agents\\Act Like Staff Engineers?}

\hypersetup{pdftitle={sudo L7: Can Coding Agents Act Like Staff Engineers?},
            pdfauthor={Nick Heiner, Ian Niebres, Nick Robin, Ryan McWhorter, Omar Alrifai, Suhaas Garre, Sushant Mehta, Edwin Chen},
            pdfsubject={NeurIPS 2026 Workshop on Interpreting Agent Behavior},
            pdflang={en-US},
            bookmarksnumbered=true}

\author{%
  Nick Heiner \qquad Ian Niebres \qquad Nick Robin \qquad Ryan McWhorter \\
  \bfseries Omar Alrifai \qquad Suhaas Garre \qquad Sushant Mehta \qquad Edwin Chen \\
  Surge AI \\
  \texttt{research@surgehq.ai}%
}

\begin{document}

\maketitle

\begin{abstract}
Coding agents can now turn a well-specified ticket into a patch that passes its tests. Experienced
engineers also notice risks the request never mentioned, question flawed premises, reuse what
already exists and say plainly what they did and did not check. We present \bench{}, a benchmark of
60 software-engineering tasks written by staff-level engineers, mostly in private production
codebases and each built around a place where a frontier model failed. A task pairs a realistic,
underspecified request with a hidden, expert-authored rubric of atomic criteria (\NCriteria{} in
all), tagged with eight behavioral dimensions from functional correctness to truthful reporting. An agentic
grader reads the agent's transcript, inspects and runs its code, and treats test output as
evidence. An attempt passes only if it meets every criterion of the two highest severities, and
more than half of these gating criteria test something other than functional correctness. On the
public leaderboard as of November 2026 (19 models, each in its developer's harness or in OpenCode,
three attempts per task), Claude Fable~5.2 passes 52.2\% of attempts, Claude Sonnet~5.5 50.6\% and Claude Opus~5.5
46.1\%. \mbox{GPT-6.1}~Sol, Claude Fable~5.1 and \mbox{GPT-6}~Astra pass 31--34\%, and the other 13 models at
most 18.9\%. In a human-label study, three candidate model graders each agreed with a software engineer on
85--86\% of criterion verdicts, about as often as two engineers agreed with each other. Cases from
the benchmark show how one unsupported claim can fail an otherwise strong attempt and illustrate two
sides of engineering judgment: flagging hidden hazards and avoiding needless work.
\end{abstract}

\section{Introduction}

Repository-level benchmarks made resolving real issues a standard test for coding agents
\citep{jimenez2024swebench,yang2024sweagent}, and newer suites add economic stakes, longer horizons,
original tasks and stronger verifiers
\citep{miserendino2025swelancer,deng2025swebenchpro,zhou2026featurebench,huang2026deepswe,merrill2026terminalbench}.
Most of them still define success as passing an executable verifier. Verifiers are hard to do
without, but they miss a class of failures that experienced engineers worry about most: code that
works and is still the wrong thing to ship. A password manager can implement every requested feature
and display saved passwords in plain text. A production database migration can change the schema
correctly while deleting records without confirmation or any way to roll back. A stock-alert
dashboard can look finished and never tell the user that every price is simulated
\citep{surge2026sudol7blog}. Audits of popular coding benchmarks also find tests that reject valid
solutions and accept incomplete ones \citep{openai2026signalnoise,openai2026verifiedaudit}. Both
problems argue for treating a verifier's output as evidence for a grader instead of as the score,
and for grading the whole attempt, from the commands the agent ran to what it finally told the
user.

A junior (L3) engineer is usually handed a well-defined problem and asked for a patch. A staff (L7)
engineer gets requests like ``make this faster'' or ``migrate this safely'', and much of their value
lies in what the ticket leaves out: the risk nobody mentioned, the side effect of the obvious fix,
the point where passing tests stops being enough. We present \bench{} \citep{surge2026sudol7blog},
a benchmark for that outer loop of deciding what to build and what not to build, as well as for the
inner loop of writing code that works. The paper makes four contributions.

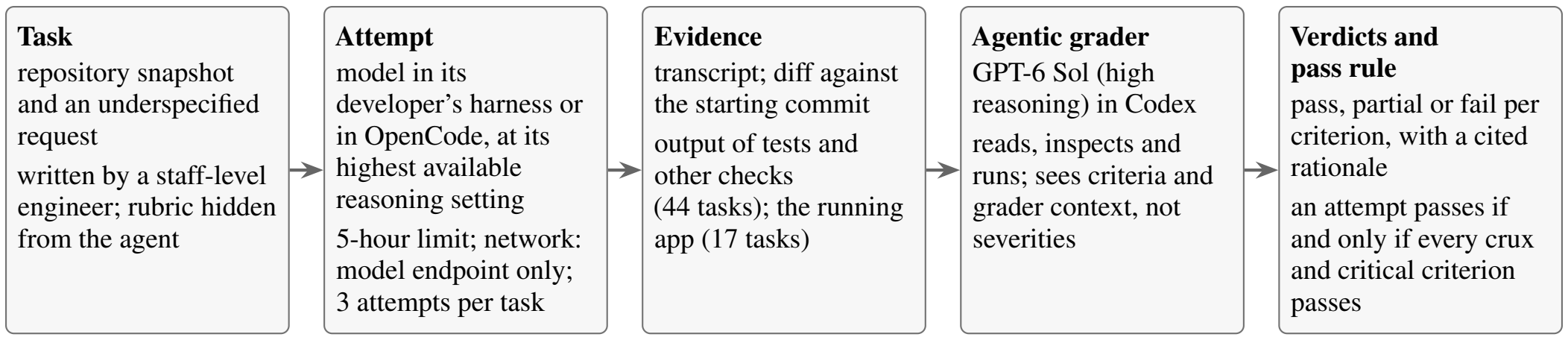
\begin{figure}[t]
  \centering
  \begin{tikzpicture}[
      box/.style={draw=black!55, rounded corners=2pt, align=flush left, text width=2.3cm,
                  inner sep=3pt, font=\scriptsize, text height=1.5ex, text depth=2.45cm,
                  fill=black!3,
                  execute at begin node={\hyphenpenalty=10000\exhyphenpenalty=10000}},
      arr/.style={-{Stealth[length=2mm]}, thick, black!60}]
    \node[box] (task) {\textbf{Task}\\[1pt]
      repository snapshot and an underspecified request\\[2pt]
      written by a staff-level engineer; rubric hidden from the agent};
    \node[box, right=0.3cm of task] (attempt) {\textbf{Attempt}\\[1pt]
      model in its developer's harness or in OpenCode, at its highest available reasoning
      setting\\[2pt]
      5-hour limit; network: model endpoint only; 3~attempts per task};
    \node[box, right=0.3cm of attempt] (evidence) {\textbf{Evidence}\\[1pt]
      transcript; diff against the starting commit\\[2pt]
      output of tests and other checks (\NCheckTasks~tasks); the running app
      (\NBrowserTasks~tasks)};
    \node[box, right=0.3cm of evidence] (grader) {\textbf{Agentic grader}\\[1pt]
      GPT-6 Sol (high reasoning) in Codex\\[2pt]
      reads, inspects and runs; sees criteria and grader context, not severities};
    \node[box, right=0.3cm of grader] (score) {\textbf{Verdicts and\\pass rule}\\[1pt]
      pass, partial or fail per criterion, with a cited rationale\\[2pt]
      an attempt passes if and only if every crux and critical criterion passes};
    \draw[arr] (task) -- (attempt);
    \draw[arr] (attempt) -- (evidence);
    \draw[arr] (evidence) -- (grader);
    \draw[arr] (grader) -- (score);
  \end{tikzpicture}
  \caption{How an attempt is graded. The grader judges each atomic criterion from the transcript,
  the code and the output of deterministic checks. Severities are applied only afterwards, and only
  criteria of the two highest severities, crux and critical, can fail an attempt.}
  \label{fig:pipeline}
\end{figure}

\begin{itemize}[leftmargin=1.5em,itemsep=1pt,topsep=2pt]
  \item \textbf{Tasks} (Section~\ref{sec:tasks}): 60 tasks written by software engineers at staff
    level or above, 38 of them in private production codebases, each built around a place where a
    frontier model failed. Not every task calls for code.
  \item \textbf{Grading} (Sections~\ref{sec:rubrics}--\ref{sec:scoring} and
    Figure~\ref{fig:pipeline}): \NCriteria{} atomic rubric criteria tagged with eight behavioral
    dimensions; an agentic grader that inspects and runs the agent's work and treats test output as
    evidence; safeguards against reward hacking; and a strict pass rule under which an attempt can
    fail on judgment alone even when its code works.
  \item \textbf{Results and interpretation} (Sections~\ref{sec:protocol}--\ref{sec:hazard}): the
    public leaderboard of 19 models from 12 developers, with its sampling uncertainty bracketed, and
    a two-sided account of engineering judgment, illustrated with cases. \emph{Hazard}
    cases reward surfacing a constraint the request did not mention; \emph{restraint} cases reward
    doing only what is needed.
  \item \textbf{Validity} (Section~\ref{sec:grader}): a human-label study in which three candidate
    graders agree with engineers on criterion verdicts about as often as engineers agree with each
    other.
\end{itemize}

\section{Related work}

\paragraph{Repository-level software engineering.}
SWE-bench \citep{jimenez2024swebench} drew its tasks from real GitHub issues, and SWE-agent
\citep{yang2024sweagent} showed that the interface between agent and computer can change results
substantially. SWE-Lancer values tasks at real
freelance payouts and includes managerial tasks in which a model chooses among implementation
proposals \citep{miserendino2025swelancer}. Other suites draw on held-out and proprietary
repositories \citep{deng2025swebenchpro}, test multi-commit feature development
\citep{zhou2026featurebench}, write original tasks with purpose-written behavioral verifiers
\citep{huang2026deepswe}, or pose hard command-line tasks with explicit constraints
\citep{merrill2026terminalbench}. One Terminal-Bench~2.0 task, for example, asks for a fastText
classifier within fixed size and accuracy limits, whereas a comparable \bench{} task would leave open
whether fastText is the right choice at all. The closest relative is FrontierCode, which asks
whether a maintainer would merge the agent's patch. Its criteria, written by maintainers of the
source repositories, cover correctness, tests, scope, style and code quality, and a patch that
misses any blocking criterion scores zero \citep{cognition2026frontiercode}. FrontierCode~2.0 puts
more weight on what its authors call thoroughness: requirements that an agent should infer from the
instructions and the existing code even when nobody states them \citep{cognition2026frontiercode2}. In both
versions the criteria judge the patch. The \bench{} grader also reads the transcript and the final
report, so an attempt whose code works can still fail for leaving a hazard unflagged, for not
questioning a flawed premise or for overstating what it checked. The right output is sometimes no
code at all.

\paragraph{Rubric-based and behavioral evaluation.}
Expert-written rubric criteria scored by a model grader underlie HealthBench
\citep{arora2025healthbench} and PaperBench \citep{starace2025paperbench}, and our earlier work
used expert rubrics as reinforcement-learning rewards \citep{mehta2026complexconstraints}.
AgentBench evaluates decision making across interactive environments \citep{liu2024agentbench},
GAIA tests tool use on real-world questions that are simple for humans \citep{mialon2024gaia}, and
our HANDBOOK.md benchmark asks whether agents can follow long policies that vary from task to task
\citep{panavas2026handbook}. Two of our recent studies characterize long-horizon agent behavior
through goal selection, verification and transfer across domains
\citep{mehta2026crossbenchmark,ritchie2026office}. \bench{} targets repository-contingent
judgment, where what counts as correct depends on facts about one codebase.

\paragraph{Graders and validity.}
LLM judges can track human preferences, but they show position, verbosity and self-preference
biases \citep{zheng2023judge,wang2024faireval,liu2023geval,panickssery2024selfpref}.
Agent-as-a-Judge extends model grading to agents that evaluate other agents, judging the whole
task-solving process rather than only its result, and applies the idea to code generation
\citep{zhuge2025agentjudge}. The \bench{} grader follows this pattern; its criteria also cover what
the request leaves unsaid, and it reads grader context written by each task's author. Whether a
score supports a conclusion depends on a validity argument: evidence that its measurement fits the
construct it is meant to capture \citep{jacobs2021measurement,raji2021everything}. Documentation is part of that argument
\citep{gebru2021datasheets}, and agentic benchmarks need explicit checks of task validity and
outcome validity \citep{zhu2025rigorousbenchmarks}. We therefore report the grading protocol, the
grader study and the evaluation conditions as part of what the scores mean.

\section{The \bench{} benchmark}

\subsection{Tasks}\label{sec:tasks}

A task is a repository snapshot plus a request of the kind engineers receive: ``Make this faster'',
``Migrate these API tokens to secure hashes'', ``Can we safely cache this?''\ or ``We have a
security audit coming up. Fix our password protection.'' Requests are underspecified on purpose, as
real tickets are. The agent has to explore the system, decide what matters, choose an approach and
verify the dangerous cases. Depending on the task, the right output is a patch, a review, an
architectural analysis, a rollout plan or an explanation of why the obvious implementation would
create a larger problem. Every task was written by a software engineer at staff level or above. The contributors include a
principal software engineer at NASA JPL, a staff engineer at Ramp, and staff engineers from Google,
Meta, Cloudera and GitHub \citep{surge2026sudol7blog}.

Most tasks come from licensed private production codebases at three companies, and the rest from
open-source projects (Table~\ref{tab:composition}). These systems carry years of accumulated
technical debt, and one of them, from a money-management app, has roughly 250,000 lines of code. The
leaderboard is public \citep{surge2026sudol7board}, and access to the task set itself can be
requested at \href{mailto:benchmarks@surgehq.ai}{\texttt{benchmarks@surgehq.ai}}. All 60 tasks are single-turn. Multi-turn tasks, which supply an
earlier conversation for the agent to resume, are excluded for now, because not every agent
harness on the board can resume a conversation it did not start.

\paragraph{Adversarial authoring and quality control.}
Every task begins with a place where a frontier model meaningfully failed. Each author is
responsible for the request, the rubric and the grader context, and checks every criterion. One
of the project's strongest engineers reviews each task. Author and reviewer iterate
together, escalate disagreements they cannot resolve to a more senior reviewer, and get feedback
from automated checks. Many attempts by several agents are then used to check whether each task is
solvable, whether its rubric covers the range of observed responses and whether its difficulty is
broad rather than one agent's edge case. The same attempts also show whether agents spread out in
performance and whether rubric-based grades order attempts the way an engineer would. Further
checks across the whole set look at the mix of repositories and task types.

\begin{table}[t]
  \caption{The \bench{} task set (60 single-turn tasks). Request style and area are tags from a
  fixed vocabulary; a task can carry several, or none.}
  \label{tab:composition}
  \centering
  \small
  \begin{tabularx}{\linewidth}{@{}lL@{}}
    \toprule
    Authors & software engineers at staff level or above \\
    Codebases & \NRepos{} repositories: 17 private production repositories from three companies
      (38~tasks) and 8~open-source projects (22~tasks) \\
    Main language & Ruby 21, TypeScript 17, Python 13, JavaScript 9, Terraform 1 (three tasks have
      two main languages; two code reviews have none) \\
    Work type & feature implementation 28, debugging and fixes 25, refactoring and migration 3,
      code~review~2, architecture and analysis 1, testing and operations 1 \\
    Request style & conversational ask 23, terse directive 17, bug report with evidence 6, question
      or consultation 5, support relay 5, ticket 2; a false premise to refute in 10 tasks \\
    Area & backend 45, frontend 27, debugging 18, domain-specific 11, security 9, reliability 9,
      long-horizon 4, architecture 3 \\
    Rubric criteria & \NCriteria{} in total; median \CritMedian{} per task (range
      \CritMin--\CritMax) \\
    Severity & crux 69, critical 305, major 632, minor 191 (the 90 extra-credit criteria have
      none); the \NGating{} crux and critical criteria gate the pass rule \\
    Category & primary intent 871 (68\%), dodged bullet 326 (25\%), extra credit 90 (7\%) \\
    Evidence & deterministic checks in \NCheckTasks{} tasks, mostly the repository's own tests (a
      linter, type checker or hidden task-specific test in some); a browser for the running app in
      \NBrowserTasks{} tasks \\
    \bottomrule
  \end{tabularx}
\end{table}

\subsection{Rubrics and the eight dimensions}\label{sec:rubrics}

Scoring by hidden unit tests would force every valid solution through one test suite, which these
open-ended tasks do not allow. Each task therefore carries an \emph{atomic rubric}: a list of
statements about the outcome, each judged on its own, which the agent never sees. Across the 60
tasks there are \NCriteria{} criteria (median \CritMedian{} per task). Each criterion has a
category, one or more behavioral dimensions and, unless it is extra credit, a severity. Nearly all
also carry an \emph{elaboration} that tells the grader how to apply them. Categories separate the
work the request needs (\emph{primary intent}, 68\% of criteria), avoiding a plausible mistake such
as breaking a neighboring flow (\emph{dodged bullet}, 25\%), and going beyond the ask, for example
by raising a real open question (\emph{extra credit}, 7\%; such criteria count only when earned). Severities are \emph{crux} (the point of the task; at most two per task, present in
\NCruxTasks{} of the 60), \emph{critical} (missing it sinks the work), \emph{major} (missing it may
sink the work) and \emph{minor} (worth getting right, rarely decisive). Crux and critical criteria
are \emph{gating}: an attempt passes only if all of them pass (Section~\ref{sec:scoring}). Each
task also ships \emph{grader context} written by its author: ground truth about the repository and
domain, with file and line references, that tells the grader what to look for.
Appendix~\ref{app:criterion} shows an illustrative criterion.

The dimensions (Table~\ref{tab:dimensions}) name the engineering behavior a criterion measures. A
criterion can carry more than one (241 do), so a single lapse, such as a final report that claims a
check that never ran, can register as both a verification miss and a truthful-reporting miss.

More than half of what the pass rule checks is not functional correctness. Of the \NGating{}
gating criteria, 203 (54\%) carry no functional-correctness tag, and 179 of those carry at least
one of verification, thought partnership, communication, truthful reporting or persistence. These
179 appear in 55 of the 60 tasks, so in almost every task an attempt whose code works can still
fail on judgment alone. For its size, truthful reporting gates most often: 66 of its 90 criteria
are gating, and they fall in 48 tasks, the same number as for functional correctness.

\begin{table}[t]
  \caption{The eight behavioral dimensions, the question each asks of an attempt, and how many
  criteria carry each tag (a criterion can carry several). Gating criteria (crux or critical)
  fail the attempt when missed; the last column counts the tasks with at least one.}
  \label{tab:dimensions}
  \centering
  \small
  \setlength{\tabcolsep}{4pt}
  \begin{tabularx}{\linewidth}{@{}lLrrr@{}}
    \toprule
    & & All & \multicolumn{2}{c}{Gating} \\
    \cmidrule(l){4-5}
    Dimension & What it asks & criteria & criteria & tasks \\
    \midrule
    Functional correctness & Does the requested implementation actually work? & 391 & 171 & 48 \\
    Engineering craft & Is the solution well designed, maintainable, reliable, secure and appropriate for the existing system? & 271 & 53 & 29 \\
    Verification & Did the agent actually check its work, including important edge cases and failure modes? & 255 & 58 & 37 \\
    Thought partnership & Does it question flawed assumptions, identify unstated risks, suggest better approaches and help the user make better decisions? & 182 & 51 & 32 \\
    Communication & Does it clearly explain its work, decisions, trade-offs and remaining limitations? & 160 & 35 & 24 \\
    Truthful reporting & Does it accurately report what it did and verified, without claiming work it never performed or concealing important failures? & 90 & 66 & 48 \\
    Persistence & Does it follow through, recover from setbacks and seek clarification when necessary? & 104 & 22 & 16 \\
    Practical judgment & Does it make sensible engineering choices, reuse existing functionality and avoid unnecessary complexity? & 106 & 14 & 7 \\
    \bottomrule
  \end{tabularx}
\end{table}

\subsection{Grading}\label{sec:grading}

When the agent finishes, a verifier runs in the agent's container (Figure~\ref{fig:pipeline}). It
captures the session transcript and the agent's changes against the task's
starting commit, so committed and uncommitted work both count. For
\NCheckTasks{} of the 60 tasks it also runs deterministic checks set for that task. These are
usually the repository's own tests, in some tasks a linter or type checker, and in some a
task-specific test the agent never sees. Failures known to predate the agent's change are flagged,
and all of this output goes to the grader as evidence, never as a score.

The grader, \mbox{GPT-6}~Sol at high reasoning effort running in Codex, then grades
\emph{agentically}: it reads the transcript, inspects and runs the code, and, where the task image
includes a browser, can load the application and take screenshots. It receives a shared grading
prompt, the task's grader context, the criteria and the check output, and returns a verdict for
every criterion (pass, partial or fail) with a rationale citing evidence such as moments in the
transcript, files and command output. The grader never sees severities; they are applied afterwards,
so stored verdicts can be re-scored under any rule without re-grading. Because every criterion keeps
its dimension tags and every verdict cites its evidence, a failed attempt can also be traced to the
behaviors behind the criteria it missed, which the leaderboard's single pass rate does not show.

\mbox{GPT-6}~Sol was chosen after an internal comparison of candidate graders that included the
human labels of Section~\ref{sec:grader} and judged it on par with frontier graders at a much lower
price. Section~\ref{sec:grader} reports the human-label study for three other candidates.
\mbox{GPT-6}~Sol is not itself on the leaderboard, and the \mbox{GPT-6.1}~Sol listed there is a
different model from the same developer.

Grading runs in the agent's own container, so the benchmark has to guard against reward hacking.
Agent and grader can reach only their inference endpoints, the agent has no history beyond its
starting commit, and the rubric is copied in only after the agent's turn ends. A doctored test
suite is visible to the grader, which sees the agent's changes alongside the test output. The
verifier also records tamper flags for every attempt, and Appendix~\ref{app:safeguards} lists the
safeguards.

\subsection{Scoring}\label{sec:scoring}

An attempt passes only if every crux and critical criterion receives a \emph{pass}. A partial
counts as a miss, and major, minor and extra-credit criteria never block. A task's \emph{pass
fraction} is the share of its attempts that pass, and a model's \emph{pass rate} is the mean of its
pass fractions over the 60 tasks. The benchmark also emits a severity-weighted continuous reward
that credits partial progress, but the leaderboard reports the pass rate. The rule is strict on
purpose. A single confident claim that the work does not support can fail a critical criterion and
take an otherwise strong attempt down with it, much as one overstated claim in a colleague's report
makes a reviewer doubt the rest.

\section{Evaluation protocol}\label{sec:protocol}

We report the public leaderboard as of November 2026 \citep{surge2026sudol7board} and take every
pass rate in this paper from it, so that each can be checked against the board.\footnote{Scores on
the current board differ from those in the benchmark's launch announcement
\citep{surge2026sudol7blog}. Since launch, about half of the requests have been reworded (mostly for
tone), the rubric and grader wording has been tidied, and the protocol has changed from one attempt
graded three times to three attempts graded once. We report only the current board.} Each of the 19
models made three independent attempts at every task, at the highest reasoning setting available
(Table~\ref{tab:leaderboard}). Models ran in their developers' own agent
harnesses where the benchmark supports one, such as Claude Code for Claude models and Codex for GPT
models, and otherwise in the open-source OpenCode harness, for example for the Kimi and GLM models. Attempts have a five-hour limit and can reach only the model's
inference endpoint. Web search and web fetch are off, and no one answers questions during an
attempt, so an agent that needs clarification can ask only in its final report, which ends the
attempt. Each attempt is graded once. An attempt that hits the five-hour limit or ends in an agent
or infrastructure error is discarded, even if it was graded, and re-run instead of being scored as
a failure. If no re-run yields a valid grade, the attempt is left out of the pass rate.

\section{Results}\label{sec:results}

\begin{table}[t]
  \caption{The \bench{} leaderboard, snapshot of 13 November 2026 \citep{surge2026sudol7board}.
  Each row is a model in its harness (Section~\ref{sec:protocol}). Pass rate (the leaderboard's ``Score''): share of attempts in which every crux and critical
  criterion passes, averaged over the 60 tasks (three attempts each). Spacing marks the score groups
  of Section~\ref{sec:results}. Reasoning: as listed on the leaderboard;
  Qwen~3.8 Max ran in Qwen Code, to which the benchmark passes no reasoning setting. Ties share a
  rank. \textsuperscript{\textdagger}This rate is not a whole number of passing attempts out of 180,
  so it was computed over fewer attempts than the others (Section~\ref{sec:protocol}); the
  leaderboard does not say how many.}
  \label{tab:leaderboard}
  \centering
  \small
  \begin{tabular}{@{}rlllrl@{}}
    \toprule
    Rank & Model & Developer & Reasoning & Pass rate (\%) & \\
    \midrule
     1 & Claude Fable 5.2   & Anthropic         & Adaptive/Max & 52.2 & \pbar{52.2} \\
     2 & Claude Sonnet 5.5  & Anthropic         & Adaptive/Max & 50.6 & \pbar{50.6} \\
     3 & Claude Opus 5.5    & Anthropic         & Adaptive/Max & 46.1 & \pbar{46.1} \\
    \addlinespace[3pt]
     4 & GPT-6.1 Sol        & OpenAI            & Max          & 33.9 & \pbar{33.9} \\
     5 & Claude Fable 5.1   & Anthropic         & Adaptive/Max & 32.2 & \pbar{32.2} \\
     6 & GPT-6 Astra        & OpenAI            & Max          & 31.1 & \pbar{31.1} \\
    \addlinespace[3pt]
     7 & MiniMax M3.1 Flash & MiniMax           & Max          & 18.9 & \pbar{18.9} \\
     8 & Kimi K3.1 Pro      & Moonshot AI       & Max          & 16.7 & \pbar{16.7} \\
     8 & Qwen 3.8 Max       & Alibaba Cloud     & xHigh        & 16.7 & \pbar{16.7} \\
    10 & Muse Wonder 1.0    & Meta              & Max          & 16.1 & \pbar{16.1} \\
    10 & Gemini 4 Argon     & Google            & High         & 16.1 & \pbar{16.1} \\
    12 & Hy4 Preview        & Tencent           & High         & 15.6 & \pbar{15.6} \\
    13 & Inkling 2 Small\textsuperscript{\textdagger} & Thinking Machines & xHigh & 14.1 & \pbar{14.1} \\
    14 & Mistral Large 4    & Mistral AI        & High         & 13.9 & \pbar{13.9} \\
    15 & Muse Spark 1.3     & Meta              & Max          & 12.8 & \pbar{12.8} \\
    16 & Grok 4.7           & xAI               & xHigh        & 12.2 & \pbar{12.2} \\
    16 & Kimi K3            & Moonshot AI       & Max          & 12.2 & \pbar{12.2} \\
    18 & Gemini 3.8 Flash   & Google            & High         & 10.6 & \pbar{10.6} \\
    19 & GLM 5.3            & Zhipu AI          & Max          &  9.4 & \pbar{9.4} \\
    \bottomrule
  \end{tabular}
\end{table}

\paragraph{No agent passes much more than half its attempts.}
Three Claude models lead the board (Table~\ref{tab:leaderboard}). Fable~5.2 passes 52.2\% of
attempts, Sonnet~5.5 50.6\% and Opus~5.5 46.1\%. \mbox{GPT-6.1}~Sol (33.9\%), Claude Fable~5.1 (32.2\%)
and \mbox{GPT-6}~Astra (31.1\%) form a second group, and the other 13 models, from ten developers, pass
between 9.4\% and 18.9\%. The benchmark can register generational progress: Fable~5.2 passes
20.0~points more than Fable~5.1, a gap that clears even the conservative interval below.

\paragraph{The large gaps hold up.}
The leaderboard publishes one pass rate per model and no per-task results. Instead of estimating the
task-level uncertainty directly, we bracket it from the published rates
(Appendix~\ref{app:uncertainty}), so that the intervals, like the pass rates, can be checked against
the board. A pass rate is a mean of
60 task-level pass fractions, and a model's three attempts on a task share that task's difficulty.
If every task were equally hard for the model and its attempts were independent, a 95\% interval at
a 50\% pass rate would be about \textpm\,7.3~points, the optimistic case. If its attempts on each
task always agreed, it would be about \textpm\,12.7~points, the conservative case. All 39 gaps
between a top-group model and one of the other 13 (at least 27.2~points) exceed even the
conservative interval for a difference between two models, as does the 20.0-point gap between
Fable~5.2 and Fable~5.1, against a conservative interval of 17.3~points. In all, 78 of the board's 171
pairs clear their conservative intervals (Appendix~\ref{app:uncertainty}).
Gaps of a few points, such as those within the top group (at most 6.1~points), are within noise.

\paragraph{Careful reporting can outweigh deeper investigation.}
The 4.5-point gap between Sonnet~5.5 and Opus~5.5 is within noise, so we do not read it as a
ranking. An internal per-task comparison of the two found that
neither dominated: each passed tasks the other failed. Opus tended to investigate more deeply,
while Sonnet's wins came mostly from caution in what it asserted. Sonnet's final reports stayed with
what it had changed and checked, and it more often confirmed that a request's premise held before
acting on it. Under a pass rule in which one missed crux or critical criterion fails the attempt, a
single unsupported claim can cost more than extra depth earns. In the fraud-labels task, for example,
the agent is asked to make the fraud labels in a training dataset trustworthy. Opus repaired more
than Sonnet, yet opened its report with ``I've fixed the fraud labeling'' while known labeling
problems remained, and failed. Sonnet described the one real bug it had fixed and the checks it had
added to expose the remaining problems, and said what it could not fix without the actual data.

\section{Interpreting failures: hazard and restraint}\label{sec:hazard}

The hazard and restraint cases below, summarized in Table~\ref{tab:cases} with the reporting case
of Section~\ref{sec:results}, illustrate the two sides of engineering judgment.

\begin{table}[t]
  \caption{Cases discussed in Sections~\ref{sec:results} and~\ref{sec:hazard}, and what the rubric
  rewards. All three tasks are among the 60 on the leaderboard.}
  \label{tab:cases}
  \centering
  \small
  \begin{tabularx}{\linewidth}{@{}lLL@{}}
    \toprule
    Kind & Case & What the rubric rewards \\
    \midrule
    Hazard & A one-line Terraform typo fix whose deployment could replace a load balancer & Noticing and flagging the replacement before anything is deployed \\
    \addlinespace[2pt]
    Restraint & Opus~5.5 built a second system for splitting donations, with 267 passing tests, although the feature existed & Pointing to the existing feature and asking what the user needs \\
    \addlinespace[2pt]
    Reporting & Opus~5.5 opened its report with ``I've fixed the fraud labeling'' while known labeling problems remained & Reporting what was fixed, what new checks expose, and what remains \\
    \bottomrule
  \end{tabularx}
\end{table}

\paragraph{Hazard.}
In a hazard case the request is plausible, but following it literally misses a real constraint, so a
strong attempt investigates and surfaces the consequence. Hazards can hide in one-line changes. In
one task, the user asks the agent to fix a typo in a Terraform variable. The fix itself is trivial,
but the rubric asks whether the agent also notices that deploying it could replace a load balancer
\citep{surge2026sudol7blog}, and whether it says so before anything is deployed. An agent that fixes
only the typo can deliver exactly what was asked and still leave the user exposed.

\paragraph{Restraint.}
Restraint cases pull the other way. The request is sound as asked, or what it asks for already
exists, so a strong attempt does what is needed and no more, without inventing a vulnerability or
building a parallel system. In a charitable-giving app, a user asked: ``Can you let people split
donations across a couple of different organizations instead, as long as it still adds up to what
they've pledged?'' Opus~5.5 built a new interface and database structure and tested them
thoroughly, with 267 tests and four browser tests passing. But the app already supported allocating
donations across organizations, and Opus had even encountered the relevant code. The result is two
code paths for one feature. A staff engineer would point to the existing functionality and ask what
the user needs, which is what the rubric checks \citep{surge2026sudol7blog}.

\paragraph{Both sides in one benchmark.}
A benchmark of hazards alone would reward an indiscriminately suspicious agent, and one of benign
requests alone would reward shallow compliance. \bench{} also includes restraint cases like the
donation task, checked through criteria such as whether the agent points out that a requested
feature already exists \citep{surge2026sudol7blog}.

The launch announcement reads the donation case as a failure to question the request, the same root
as a missed hazard \citep{surge2026sudol7blog}. Opus had encountered the existing code, which
suggests that the lapse came after the search, in understanding that code or in judging the request.
Two readings of the link between hazard and restraint failures fit this case, and they differ on
restraint tasks that penalize needless caution. If
one habit, accepting a request unexamined, drives both kinds of failure, results on those tasks
should not fall as hazard results rise. If agents instead trade scrutiny against compliance, those
results should fall. Per-task results with hazard and restraint labels, compared at a fixed overall strength, would
tell the two apart.

\section{Grader validity}\label{sec:grader}

\begin{table}[t]
  \caption{Agreement with a reference software engineer on the five tasks that a second engineer
  also labeled, drawn from the benchmark's wider task pool. The three model graders are candidates;
  the production grader, \mbox{GPT-6}~Sol, is not among them. Criterion verdicts: share of 293
  verdicts (one criterion judged on one attempt) on which each rater agreed with the reference
  engineer, with 95\% confidence intervals as reported by the study. Preferred attempt: pairs of
  attempts at the same task on which both preferred the same attempt, out of 15.}
  \label{tab:grader}
  \centering
  \small
  \begin{tabular}{@{}lcccc@{}}
    \toprule
    & Second engineer & Gemini 4 Argon & Claude Fable 5.1 & GPT-6 Astra \\
    \midrule
    Criterion verdicts (293) & 85\% (78--92) & 85\% (79--90) & 86\% (80--91) & 86\% (78--93) \\
    Preferred attempt (15 pairs) & 14 of 15 & 14 of 15 & 13 of 15 & 9 of 15 \\
    \bottomrule
  \end{tabular}
\end{table}

Candidate model graders were compared with software engineers on labeled attempts at tasks from the
benchmark's wider task pool. Engineers judged individual rubric criteria and said which of two
attempts at the same task they preferred. On five of the study's 14 tasks, a second engineer repeated
the labeling without seeing the first engineer's labels, which gives a human--human baseline. Where
the model graders were unanimous, both engineers saw that verdict before confirming or overturning
it. Every figure in Table~\ref{tab:grader} comes from those five tasks. On criterion verdicts, each
model grader agreed with the reference engineer about as often as the second engineer did, 85--86\%
against 85\%. On the better of two attempts, Gemini~4~Argon matched the reference engineer on 14 of
15 pairs,\footnote{With 15 pairs, one pair moves agreement by almost 7~points, and an exact 95\%
binomial interval for 9 of 15 runs from 32\% to 84\%.} as often as the second engineer did. Claude
Fable~5.1 matched on 13 and \mbox{GPT-6}~Astra on 9.

The production grader comes from the same developer as two of the evaluated models,
\mbox{GPT-6.1}~Sol and \mbox{GPT-6}~Astra. LLM judges can favor their own generations
\citep{panickssery2024selfpref}, but both models rank below the top three Claude models, so any such
preference did not decide the top of the board.

\section{Conclusion}

As measured by \bench{}, coding agents do not yet reliably act like staff engineers: even the
strongest pass only about half their attempts. More than half of the gating criteria test something
other than functional correctness, and almost every task has at least one of them, so an attempt with
working code can still fail on an unflagged risk or one overstated claim. Verdicts are stored per
criterion, each with its criterion's dimension tags, so the same attempts can be re-scored under
other rules and read behavior by behavior. An L3 is handed a well-defined problem and asked for a pull request; an L7
is handed responsibility. \bench{} asks whether coding agents are ready for more of it.

\renewcommand{\bibfont}{\small\raggedright}
\bibliographystyle{plainnat}
\bibliography{references}

\clearpage
\appendix
\section{An illustrative rubric criterion}\label{app:criterion}

To protect the benchmark, we do not publish real criteria. The criterion below was written for
this paper, in the benchmark's format, for a hypothetical request to acknowledge failed incoming
webhooks with HTTP 200 so that partners stop retrying them.

\begin{quote}\small
\textbf{Criterion:} The response should either ensure that a webhook event whose processing failed
is not lost (for example, by storing it durably before acknowledging it) or, if it acknowledges
failed events as the user asked, warn the user that this can lose them permanently.\\
\textbf{Category:} dodged bullet. \quad \textbf{Severity:} critical.\\
\textbf{Dimensions:} thought partnership, communication.\\
\textbf{Elaboration for the grader:} Any design that stops a failed event from being lost
qualifies: persisting the payload before returning 200, an internal retry queue, or keeping non-2xx
responses for failures. Judge the outcome, not the route. Fails when the agent returns 200 for
failed events with no durable copy and does not flag the consequence.
\end{quote}

\section{Safeguards against reward hacking}\label{app:safeguards}

Tasks run in Harbor-format containers. The safeguards are:
\begin{itemize}[leftmargin=1.5em,itemsep=1pt,topsep=2pt]
  \item \textbf{Network.} The agent phase and the grading phase can reach only their inference
    endpoints.
  \item \textbf{No answer to read.} The agent has no repository history beyond its starting
    state, and the rubric is copied into the container only after the agent's turn ends.
  \item \textbf{Checks are evidence.} Test and lint output goes to the grader alongside the
    changes, so a rewritten or short-circuited test suite is visible; output crafted to look like
    part of the grading prompt is neutralized by context tags.
  \item \textbf{Leftovers.} Processes the agent leaves running are stopped before grading, after
    the deterministic checks and after the grader finishes; instructions, skills, hooks and
    agent-context files (such as \texttt{AGENTS.md} or \texttt{CLAUDE.md}) left in the workspace
    are ignored by the grader.
  \item \textbf{Change capture.} The step that captures the agent's changes does not trust the
    repository's git configuration, so hooks or ignore rules cannot hide an edit; grading tools run
    isolated from the workspace.
  \item \textbf{Forged grades.} The grade is taken from the grader's own output, not from a file
    the agent could modify, and the grader binary's hash is checked against its release.
  \item \textbf{Tamper flags.} For each attempt, the verifier records suspicious state, such as a
    modified grader binary, altered system packages, a pre-existing reward file, git configuration
    that rewrites diffs, or a starting commit dated after the container started.
\end{itemize}

\section{Bracketing the sampling uncertainty}\label{app:uncertainty}

Let $x_t\in\{0,1/3,2/3,1\}$ be a model's pass fraction on task $t$ and $\hat p=\sum_t x_t/60$ its
pass rate. The task-bootstrap variance of $\hat p$ is about $\mathrm{var}(x_t)/60$. If $\rho$ is
the correlation between two of the model's attempts on the same task,
$\mathrm{var}(x_t)\approx\hat p(1-\hat p)(1+2\rho)/3$. With attempts independent given the task's
difficulty, $0\le\rho\le1$: $\rho=0$ when all tasks share one difficulty, and $\rho=1$ when every
task is all-or-nothing. The 95\% half-width therefore lies between
\[
  1.96\sqrt{\hat p(1-\hat p)/180}\qquad\text{and}\qquad 1.96\sqrt{\hat p(1-\hat p)/60},
\]
that is, 7.3 and 12.7~points at $\hat p=0.5$. For a difference between two models we use the
corresponding unpaired values,
\[
  1.96\sqrt{\bigl(\hat p_1(1-\hat p_1)+\hat p_2(1-\hat p_2)\bigr)/n},
\]
with $n=180$ (optimistic) and $n=60$ (conservative). Both treat the 60 tasks as independent, as a
task bootstrap does. Both also ignore the positive correlation that shared tasks usually induce, so
a paired task-level interval, which needs per-task results, could be narrower than either.

The largest gap within the top group, 6.1~points between Fable~5.2 (52.2\%) and Opus~5.5 (46.1\%),
is below even the optimistic value (10.3~points). The 12.2-point gaps between Opus~5.5 and
\mbox{GPT-6.1}~Sol (33.9\%) and between \mbox{GPT-6}~Astra (31.1\%) and MiniMax~M3.1~Flash (18.9\%) exceed their
optimistic values (10.0 and 8.9~points) but not their conservative ones (17.4 and 15.3~points). The
27.2-point gap between Opus~5.5 and MiniMax~M3.1~Flash and the 20.0-point gap between Fable~5.2
and Fable~5.1 (32.2\%) exceed even their conservative values (16.0 and 17.3~points). Across all 39
pairs of a top-group model and one of the other 13, gaps are at least 27.2~points and conservative
values at most 16.1~points (Sonnet~5.5 against MiniMax~M3.1~Flash).

Of the board's 171 pairs, 78 exceed their conservative value: these 39, 34 of the 39 between the
second group and the other 13, and 5 of the 9 between the top and second groups (Fable~5.2 against
all three; Sonnet~5.5 against Fable~5.1 and \mbox{GPT-6}~Astra). None of the 84 pairs within a group does.
For example, the 4.5-point gap between Kimi~K3.1~Pro (16.7\%) and Kimi~K3 (12.2\%) is below even
its optimistic value (7.2~points). Gaps here and in Section~\ref{sec:results} are differences of the published
one-decimal rates. Every rate except Inkling~2~Small's matches a whole number of passing attempts
out of 180 (Table~\ref{tab:leaderboard}). Computed from those counts, a gap can differ by up to 0.1~point
(both 4.5-point gaps are 8 attempts, or 4.4~points), and no comparison among those rows changes.

\end{document}